# Centre-finding in *E. coli* and the role of mathematical modelling: past, present and future

Seán M. Murray[1] and Martin Howard[2]

[1] Max Planck Institute for Terrestrial Microbiology & LOEWE Centre for Synthetic Microbiology (SYNMIKRO), Karl-von-Frisch Strasse 16, 35043 Marburg, Germany

[2] Computational and Systems Biology, John Innes Centre, Norwich Research Park, Norwich, NR4 7UH, United Kingdom

Correspondence to:

Seán M. Murray: sean.murray@synmikro.mpi-marburg.mpg.de

Martin Howard: martin.howard@jic.ac.uk

## Abstract

We review the key role played by mathematical modelling in elucidating two centre-finding patterning systems in *E. coli*: midcell division positioning by the MinCDE system and DNA partitioning by the ParABS system. We focus particularly on how, despite much experimental effort, these systems were simply too complex to unravel by experiments alone, and instead required key injections of quantitative, mathematical thinking. We conclude the review by analysing the frequency of modelling approaches in microbiology over time. We find that while such methods are increasing in popularity, they are still probably heavily under-utilised for optimal progress on complex biological questions.

## Introduction

Interdisciplinary research is of increasing importance in understanding complex natural phenomena. Mathematical modelling in biology is a central example of such interdisciplinary work, and has a well-established tradition going back many decades. However, it is probably fair to say that the promise of modelling in biology has not yet been fully exploited. In this review we examine two cases where mathematical modelling has played a key role in shaping our biological understanding: division site positioning and DNA partitioning, analysing the lessons these two cases hold for interdisciplinary modelling approaches in biology more generally.

Many rod-shaped bacteria roughly double in length before dividing symmetrically at the cell centre to form two equal-sized daughter cells. How this central division site is specified has been a central question in microbiology for many decades. Experimentally, in the last decade of the last century, this process in *E. coli* was shown to be governed by the MinCDE proteins. Remarkably, these proteins were found to oscillate from pole-to-pole inside a single cell. In the first part of this review, we examine how mathematical modelling together with further experiments in the early 2000s finally unlocked the basis of the oscillations and thus of division site positioning in *E. coli*. This part of the review is more historical in nature and we refer to more recent reviews to cover subsequent developments. Note that our historical view is not meant to imply that the Min work is only of historical interest. In fact, quite to the contrary, the Min system is still a very active area of research. Here, we merely review how mathematical modelling provided critical early impetus to the field.

In the second part of the review, we move on to examine a closely-related problem, how DNA, and especially low copy number plasmids, are positioned to ensure stable inheritance. This has been a topic of recent intense interest and we review how a recent theoretical idea, flux-balance, together with experiments, has again made a central contribution.

Both these cases are examples of fruitful interaction between theory and experiment on systems too complex to be unpicked by experiments alone. Both illustrate how mathematical modelling can, when judiciously employed, provide a step change in our ability to extract mechanistic understanding. We then conclude this review by providing a perspective on the role and future potential for mathematical modelling in microbiology. In our opinion such approaches are still severely under-utilised, a bottleneck which we believe is holding back faster progress in understanding complex biological processes.

## Historical centre finding: cell division positioning in *E. coli* mediated by the oscillatory MinCDE system

The first decisive progress in understanding midcell division specification in *E. coli* was the identification, more than half a century ago, of a strain that produced so-called "minicells" [1]. Minicells are abnormally small cells without chromosomes that arise from aberrant polar, rather than midcell, division events. Molecular insight into how these minicells arose took many years to achieve, with the decisive breakthrough not occurring until the late 1980s, with the characterisation of the *minB* locus [2]. This operon, encoding three genes, *minC*, *minD* and *minE,* was shown to be responsible for the minicell phenotype. The molecular roles of these three genes then took a further decade of work to unravel. MinC was shown to be an inhibitor of FtsZ [3], a key protein in the formation of the cytokinetic ring necessary for dividing the cell in two, with MinC found to partner the MinD protein [4]. Without the third protein, MinE, this combination of proteins was found to inhibit cell division at all locations. In the presence of MinE, however, this inhibition was relaxed specifically at midcell, allowing symmetric division. MinE was thus referred to as conferring topological specificity on the system [2]. Quite how this was achieved still remained opaque.

A further key advance was achieved through live, fluorescent protein fusions, which for the first time laid bare the remarkable spatiotemporal dynamics of the Min proteins. MinD [5] and MinC [4,6], and later MinE [7], were found to undergo spatiotemporal oscillations from pole to pole inside single cells. The oscillations occur on a timescale of minutes, much faster than even the fastest cell cycle time. First MinC and MinD assemble on the membrane in one half of the cell, followed by the formation of the MinE ring around the cell periphery at midcell. This MinE ring then moves towards the pole, apparently displacing the MinC and MinD which tend to reassemble in the opposite cell half. Once the MinE ring reaches the cell pole, the MinE ring disassembles before reforming again at the cell centre before sweeping towards the other pole. In this way the pattern returns to its initial state, before the whole cycle repeats. It was also found that the presence of MinC, although essential for functionality due to its role in repressing FtsZ, was actually dispensable for the oscillations [5]. This simplification, requiring only two proteins to be explicitly modelled (MinD and MinE), was exploited by much of the subsequent mathematical modelling. Overall, the Min dynamics were perhaps the most striking example of spatiotemporal compartmentalization in bacteria, and provided a further decisive blow against the previously widely-held view of bacteria as mere bags of enzymes.

Further progress was also made on the biochemistry of the Min system. Specifically MinD was shown to be an ATPase [8] whose localization was ATP-dependent: on the membrane when bound to ATP, but in the cytoplasm when in an ADP form [9]. Moreover, formation of the MinE ring was found to depend on MinD [10], with MinE then stimulating MinD ATPase activity [9,11]. This process causes MinD membrane dissociation, with nucleotide reloading by MinD then occurring in the cytoplasm, permitting membrane rebinding (Figure 1).

Through the spatiotemporal dynamics of the Min proteins, it was already clear by around 2000 that midcell division specification was achieved through a time-averaged minimum in the MinC concentration [7]. Because of the spatiotemporal oscillations, this time-averaged minimum occurred at midcell. Polar division events were repressed by higher MinC concentrations, with off-centre divisions additionally repressed by the nucleoid, through the so-called nucleoid occlusion mechanism, later shown to be mediated by the nucleoid-associated protein SlmA [12]. How the Min oscillations arose remained, however, a profound mystery. One popular idea invoked the existence of topological markers that would mark midcell and somehow orchestrate the Min dynamics so that the MinCD proteins would avoid this location [13]. Quite how these marker proteins would themselves find the centre was unclear. This explanation did not, therefore, resolve the problem but merely pushed the same spatial targeting problem onto a different set of hypothetical proteins.

This was the state of affairs around 2001: much was known about the Min proteins, their interactions and biochemistry, but it was not at all clear how these elements combined to generate spatiotemporal oscillations. At this point, three mathematical modelling papers were published within a few months of each other in 2001 and 2002: by Meinhardt and de Boer [14]; Howard, Rutenberg and de Vet [15]; and Kruse [16], respectively. Despite significant differences, these models all proposed variants of the same key idea, namely that the interactions between the Min proteins, and their diffusion, spontaneously generate the spatiotemporal oscillations without the need for any other hypothetical factors. In this way, the Min oscillations are an emergent, self-organising phenomenon, reliant on an underlying dynamical instability. This instability causes a homogenous distribution of proteins to be unstable via an amplification of any small concentration fluctuations, eventually leading to pole-to-pole oscillations. The three models were also technically implemented in a similar way, using a deterministic partial differential equation approach. One prime motivating experiment for these models was the behaviour of the Min proteins in filamentous cells, where division had been inhibited. In this case, it was seen that the oscillatory pattern broke up into multiple bands inside single filamentous cells [5]. Such a pattern exhibits the hallmark of a characteristic wavelength, a phenomenon that was already well established from patterning reaction-diffusion systems in development. This connection made it reasonable to suppose that the Min dynamics might have a similar underlying cause.

The Meinhardt model is a pure reaction-diffusion system, where the instability is generated by a mechanism of short-range activation and long-range substrate depletion. This model builds on a long line of activator-inhibitor models stretching back to Meinhardt and Gierer [17], implementing a special case of a Turing instability. A key part of activator/inhibitor models is an activator whose protein production is enhanced autocatalytically, but which diffuses slowly. The activator is balanced by the second key component, an inhibitor, also produced by the activator, but which diffuses more rapidly and thus disperses to inhibit the activator. In an activator-depletion mechanism, the role of the inhibitor is replaced by a rapidly diffusing substrate component, which is depleted by autocatalytic conversion into the activator. In the Meinhardt Min model, this slowly-diffusing activator is membrane MinD, with the substrate being much more rapidly diffusing cytoplasmic MinD. The final ingredient to generate oscillations is the presence of a local inhibitor (MinE), stimulated by the activator (membrane MinD), whose accumulation causes a breakdown of activator production. In regions of parameter space, such a model can indeed spontaneously generate spatiotemporal oscillatory behaviour, as well as the multiple oscillatory bands of filamentous cells [14].

In order for the model to generate long-term dynamics, production of MinD in the cytoplasm is required to compensate for degradation of the activator on the membrane. In the absence of protein production, the dynamics would quickly die out due to Min protein degradation on fast (second) timescales, leaving the cell devoid of any Min proteins. As well as being extremely costly in energetic terms, such a mechanism was actually already ruled out experimentally even in 2001. Earlier experiments had already shown that the oscillations were persistent over many oscillation cycles even when cells were treated with chloramphenicol to inhibit protein synthesis [5]. Hence, despite the model's success in reproducing the Min protein oscillatory phenomenology, a different type of model was clearly required.

The Howard et al and Kruse models took such a different approach by abandoning the traditional activator/inhibitor/depletion models and instead developing models fundamentally based on conservation of protein number. In these models, protein spatiotemporal oscillations occur via relocation of existing proteins without any production or degradation, processes which were assumed to be unimportant on the short timescale of the oscillations. Both models were thus "mass conserving", taking note from the chloramphenicol experiments, as well as being much less resource intensive for the cell. The drive behind the out of thermodynamic equilibrium oscillations was therefore not due to protein synthesis/degradation as in the Meinhardt model, but rather to MinE stimulated ATP hydrolysis and subsequent nucleotide exchange for MinD. Both models are again able to recapitulate the observed spatiotemporal oscillations both in normal sized and filamentous cells [15,16].

However, there are still fundamental differences between the Howard et al and Kruse models. The model of Kruse, as well as incorporating cytoplasmic/membrane unbinding/binding reactions and diffusion also requires the MinD proteins to display self-aggregating behaviour once bound to the membrane. Such behaviour has not been decisively observed experimentally *in vivo*, though it is difficult to positively identify. By contrast the model of Howard et al uses only reaction-diffusion terms to successfully reproduce the Min phenomenology. Although the reaction kinetics have subsequently been greatly refined (see, e.g. [18]), the model did for the first time conclusively show that a mass-conserving pure reaction-diffusion model can generate patterning dynamics. The vast majority of succeeding models of the Min system have built on this foundation to construct more elaborate and more realistic models of this type.

In the period immediately following the introduction of these models, further progress was made by adding two main features: more realistic geometries or stochastic effects [19] (or both combined [20,21]). For the former, the first generation of Min models had been mostly 1d in nature and thus only accounting for the true cylindrical geometry of *E. coli* cells in a crude way. Subsequent models injected much more accurate 3d geometries capable of generating more realistic Min protein dynamics, but without fundamentally altering the underlying mechanisms. Stochastic effects turned out to be a similar story: the Min proteins are expressed in wild-type cells at copy numbers high enough (~1000) [22] to diminish low copy number fluctuation effects. In effect the stochastic models again reproduced the phenomenology of the deterministic models. However, the Min system did form an effective arena for the development and testing of spatiotemporal stochastic reaction-diffusion models (e.g. Smoldyn [23], MesoRD [24]) that have proved extremely useful for other cell biology applications.

Perhaps the central prediction of the early Min modelling was that the spatiotemporal oscillations were an emergent, self-organised phenomenon requiring only the Min proteins themselves, ATP and membrane. This prediction was tested in spectacular fashion in subsequent *in vitro* experiments [25], which put together purified MinD and MinE, together with ATP and a supported lipid bilayer. Remarkably, this *in vitro* combination generated spontaneous Min protein patterning involving travelling waves and spiral structures. While there were clear differences with the *in vivo* dynamics (for example, the length scales were much longer), these experiments formed a powerful conformation of the modelling predictions [25]. Subsequent work has refined the *in vitro* methodology leading to more realistic dynamics closer to those observed *in vivo*, including oscillations. The quest for more realistic theoretical models has also continued, with continued refinement of reaction kinetics. We refer to a series of recent excellent review articles that discuss these developments in more detail [26–28].

Looking back on the early Min modelling effort with the benefit of nearly two decades hindsight, several observations stand out. The first is that the models injected a radically new idea into subcellular protein dynamics, an idea (self-organised protein patterning) that previously had only been applied at much longer, developmental length scales. Furthermore, the Kruse and Howard et al models utilised a mass-conserving implementation that was new, to our knowledge, in biology. Second, this concept was essential for explaining the phenomenon, but could not have been contributed by the experimental community, where such theoretical thinking was not (and still is not) widespread. No amount of further biochemical characterisation or fluorescence imaging experiments would have revealed the underlying mechanism. These experiments were essential for exposing the building blocks of the Min dynamics but on their own could not have revealed how these ingredients combine to self-organise the oscillations. Third, the theoretical ideas were contributed at an auspicious time: a great deal was already known about the Min dynamics by 2001, enough to construct a plausible theory. Of course, there are many ways in which *E. coli* could have used to find its centre, but evolution arrived at one possibility. Had the theory proceeded in isolation, other mechanisms might have been elaborated that bore no relation to the mechanism actually implemented. For example, oppositely directed, stationary concentration gradients could also be used as a centre-finding mechanism and are perhaps the most obvious method. However, while such a system does appear to be implemented in *B. subtilis*, *E. coli* selected another way, spatiotemporal oscillations, that would have been completely implausible to conjecture prior to the emergence of the experimental data.

We next turn to a closely related centre finding problem, this time in DNA partitioning systems, that has recently been the subject of intense investigation. We will again examine the role played by modelling approaches, comparing and contrasting with the Min system.

## Contemporary centre-finding

### ParABS-mediated DNA partitioning in *E. coli*

The accurate transfer of genetic material from a cell to its progeny is a critical property for all cellular life. While eukaryotes employ a well understood machine, the mitotic spindle, the mechanisms underlying DNA segregation in prokaryotes are comparatively poorly understood. This situation has improved substantially over the last decade or so, especially for the segregation of low-copy number bacterial plasmids. In this case, unlike for high-copy number plasmids, random diffusion is not a reliable means of partitioning genetic material to daughter cells as the probability of a daughter receiving no plasmids is unacceptably high. Thus, low-copy number plasmids require an active mechanism to ensure their stable inheritance.

The process is mediated by a three-component partitioning system, called *par*, that consists of an NTPase, a DNA binding protein and a centromeric-like site [29,30]. The most well understood subfamily are type II *par* systems, defined by actin-like ATPases and typified by the ParMRC system of the R1 plasmid [31]. The actin homolog ParM forms dynamically unstable filaments that push plasmids to the cell poles, in a process not dissimilar to spindle-based mitosis. Type III systems encode a tubulin-like protein TubZ that forms stable cytoskeleton-like filaments that again push plasmids to the cell poles.

In the following, however, we focus on so-called type I systems, defined by a deviant Walker ATPase. An important distinction of this type is that the ATPase, generically termed ParA, binds DNA non-specifically and thereby uses the nucleoid as a scaffold for plasmid positioning. Secondly, rather than push plasmids to the poles, type I systems position them equally and symmetrically along the long axis of the cell. In cells with a single plasmid, the plasmid is positioned to mid-cell, thereby implementing centre-finding. In cells with two plasmids, the plasmids are positioned at the one-quarter and three-quarter positions, and so on for higher copy numbers (at least up to 5) in a regular positioning pattern[1] [32–34]. This class of proteins is also involved in the regular positioning of other low copy cellular components such as chemotactic clusters [35] and carboxysomes [36]. Indeed, ParA ATPases are the most common of the three types of partitioning system and, in particular, are the only ones found on bacterial chromosomes [37,38].

Consistent with its similarity to MinD, an early study found that ParA from the *par2* locus of the *E. coli* plasmid pB171 oscillates between the two poles of the cell with a periodicity of about 20 min [39]. This was found to be dependent on the DNA-binding protein ParB and the *parC* centromere-like site. A later study suggested that ParA forms filaments *in vivo* [40] and this was subsequently confirmed *in vitro* using electron microscopy [32]. It was then discovered that ParA dynamics and plasmid movement are intertwined. The partition complex (ParB bound centromere-like site) stimulates disassembly of the ParA structure, with the plasmid moving along with, as if pulled by, the retracting structure [33]. These dynamics were consistent with studies of the F plasmid in *E. coli* and its *par* operon *sopABC* (for clarity, we will henceforth refer to all *par* operons as *parABS*). ParA was found to form ATP-dependent filaments *in vitro* [41], exhibits oscillations *in vivo* [41,42] and to have an apparent helical structure [42,43]. Later work has studied the chromosomal Par system of *Caulobacter crescentus*, which also displays filamentous structures *in vitro* [44], providing further details for how the partition complex might be pulled by a retracting filament bundle.

However, in spite of frequent *in vitro* observations of ParA filament formation, it was not clear whether ParA forms filaments *in vivo*, especially given that the distribution of ParA more often resembled a cloud than a filament. In particular, the ParABS system of the *E. coli* P1 plasmid displayed different dynamics to that of the pB171 and F plasmids. P1 ParA forms a largely diffuse cloud on the nucleoid with weak regularly positioned foci that blink upon segregation [45,46]. Nevertheless, the system is able to regularly position plasmids [45,46]. Biochemical studies showed that ATP-bound dimers of P1 ParA undergo a slow conformational change to a state competent for non-specific DNA binding [47]. This slow exchange introduces a time delay that ensures the cytosolic pool of DNA-binding-competent ParA is well mixed within the cell, perhaps explaining the observed diffuse cloud. Furthermore, *in vitro* reconstitutions of the F [48] and P1 [49] plasmid systems featuring plasmids on an immobilized DNA carpet, revealed no evidence of filaments in either system, nor in earlier biochemical assays [47]. Consistent with that, ParA and ParB showed free exchange with the DNA carpet. The ability of ParA to induce plasmid movement was subsequently shown *in vitro* using *parS* coated magnetic beads and an external magnetic field to confine F plasmid partition complexes to a ParA-coated DNA carpet [50]. Strikingly, the authors observed directed motion of beads across the DNA coated surface, with the bead leaving a transient depletion zone in its wake. They suggested that, *in vivo,* plasmids may be confined to the narrow cytosolic space between the nucleoid and cell membrane. However, it was later shown that the partition complexes of both the F plasmid and in *Bacillus subtilis* are positioned and segregated within the volume of the nucleoid [51].

---

[1] In the case of *n* plasmids, their relative positions are $\left(i - \frac{1}{2}\right)\frac{1}{n}$ for $i = 1, 2, \ldots, n$.

Similar to the situation for the Min system, much detailed experimental knowledge of the ParABS dynamics was available. However, it remained completely unclear how the various elements combined to generate the sophisticated ParA spatiotemporal patterns. Once again, as in the Min system, mathematical modelling was needed to provide key insights to help unlock the underlying mechanisms. Early efforts in this direction centred around Min-like reaction-diffusion mechanisms [43,52]. Later work focused on how the formation of retracting ParA filaments might generate patterning and plasmid positioning [33], through a filament-length dependent detachment probability. Other related work focused on details of the ParA/ParB dynamics at the filament tip via polymer simulations [53], while another model incorporated random diffusion of the partition complex along the retracting filament bundle [54]. However, experiments casting doubt on the formation of (long) ParA filaments suggested that these scenarios might not be the whole story.

The field received critical new insights from modelling work by Sugawara and Kaneko [55] and Ietswaart et al. [34] for how ParA could generate self-organised patterning. In particular, Ietswaart et al. argued that independent of the precise ParA-mediated molecular mechanism for plasmid movement, regular positioning can be explained by the competition between ParA on either side of each plasmid. Assuming that ParA can bind anywhere along the nucleoid, and treating plasmids as sinks for ParA-ATP, they showed that the only configuration where the fluxes of ParA-ATP into each plasmid from either side balance is one where the plasmids are regularly positioned. For the simplest case of a single plasmid, if the plasmid is off-centre, then the amount of ParA binding on one side will be greater than the other. As the plasmid is a sink for ParA, this creates a steeper gradient of ParA concentration on one side as compared to the other (Figure 2a). Assuming that the plasmid is able to move up a ParA concentration gradient, consistent with experiments, then because this gradient is steeper on one side rather than the other, the plasmid will be repositioned back to the centre of the nucleoid. The model was demonstrated to form appropriate patterns analytically for a completely non-polymerising case, and then shown also to work via stochastic simulations for the case of short and long ParA filaments. The flux-balance model therefore potentially provides a unifying explanation for the varied ParA patterning dynamics observed experimentally.

The flux-balance mechanism has two important requirements. The first is that binding of ParA-ATP to the nucleoid is homogenous, therefore requiring that ParA-ATP in the cytosol is well-mixed. This is consistent with the slow conformational change that Par-ATP dimers undergo before they are competent to bind DNA [47]. Secondly, if the basal (ParB-independent) hydrolysis rate is too great, or the diffusion rate too slow, then each ParA-ATP dimer explores only a fraction of the nucleoid during a single binding event and as a result the plasmid is not able to 'sense' the entire nucleoid (Figure 2b). Only when a complex gets sufficiently close to a cell pole or another complex, is there an imbalance in the ParA fluxes to bias its movement. Combined with the persistent nature of the motion, this provides a natural explanation for the lack of positioning observed when hydrolysis is too fast, or on-nucleoid diffusion is too slow or absent, and the onset of oscillatory behaviour [56–59]. As these requirements are quite general, flux-balance has been argued to underlie positioning in other systems, e.g. PomXYZ and MukBEF (see below).

An essential requirement for plasmid positioning via ParABS dynamics is that plasmids move up the (likely non-filamentous) ParA concentration gradient. The first explanation for how this would work was theoretical. Sugawara and Kaneko [55] used thermodynamic arguments to argue that binding of ParA-ATP can generate a 'chemophoretic' pulling force on the plasmid (with flux-balance subsequently resulting in regular positioning, as described above). This mechanism was later used by Vecchiarelli *et al* [60] to explain their *in vitro* observations of directed motion of *parS* coated beads and put their earlier conceptual 'diffusion-ratchet' model [47] on a mathematical footing.

More recent studies have attempted to provide molecular explanations for force generation and have extensively employed mathematical modelling approaches. Although we will not address this question in depth, we briefly mention that two mechanisms have been proposed. The first is the DNA-relay model [59,61], based on the observation that genetic loci undergo elastic fluctuations in their position, independent of chromosome segregation [62]. The second is the Brownian-ratchet model [56,63], where the entire tether from the partition complex to the nucleoid DNA (via ParB and ParA-ATP) is considered as an elastic spring that can generate forces on the complex in a similar way to the DNA-relay model. While there has been some discussion about the differences between the two models, and about the validity of the underlying assumptions and parameters values [64], the two mechanisms for directed movement are conceptually similar. However, it is not yet resolved which one might be closer to reality.

Intriguingly, the underlying molecular dynamics for the ParABS and MinCDE systems are fundamentally rather similar. Both involve an ATPase (MinD, ParA) whose activity is stimulated by an ATPase activating protein (MinE, ParB), with the location of the ATPase switching between cytoplasm and membrane (MinD) or between cytoplasm and nucleoid (ParA). However, a fundamental difference arises from the positioning and dynamics of the ATPase activator. For the Min system, in the form of MinE, this is a diffusible molecule free in principle to occupy any site on the membrane with available MinD. For the plasmid system the equivalent protein ParB binds to the *parC* centromere-like site and is therefore not free to occupy any position on the nucleoid. Hence, for the plasmid case, the sink of ParA is at discrete locations (the plasmids), whereas for the Min system, the sink is itself a self-organised, dynamic object (the MinE ring). However, we emphasise that there are still many similarities between the two systems. Plasmid systems are also capable, at least theoretically, of displaying oscillations [56,58,59] and both are examples of mass-conserving patterning dynamics. A related system, PomXYZ of *Myxococcus xanthus,* sits between the two. It functions like the Min system to define the future site of cell division but does so in a manner very similar to plasmids [57,65].

**Self-positioning of MukBEF**

One interesting question that arises from the analysis of ParABS is whether regular positioning systems exist that consist purely of dynamic proteins rather than fixed objects like plasmids. The Min dynamics in filamentous cells arguably constitute such a system, though such filamentation does not normally arise and requires an *ftsZ* mutant, for example. One example system that has recently been investigated is that of bacterial Structural Maintenance of Chromosomes (SMC) complex [66]. SMCs are ubiquitous protein complexes involved in chromosome organisation and segregation. MukBEF, the *E. coli* SMC, forms foci at the middle and quarter cell positions in cells with one or two foci respectively, in a pattern similar to that of low copy number plasmids. These foci are dynamic, fluctuate in position and intensity, and turnover on a timescale of about 1 min. [67]. Furthermore, MukBEF has multiple conformational and stoichiometric states. Foci consist of a slowly diffusing DNA-associated state while another rapidly diffusing state is distributed uniformly within the cell [67]. Combined with the biochemical evidence for cooperative interactions, it was proposed that MukBEF foci form by a diffusion-driven instability, similar to the Min system, but with a classic stationary Turing pattern, rather than an oscillation as the output pattern [66].

However, an obstacle to the application of Turing patterning in finite systems is that model parameters typically need be fine-tuned to robustly obtain the desired wavelength, independent of initial conditions. Furthermore, even with fine-tuning, the polarity is not fixed, so that the concentration profile can equally well have a peak or a valley in the middle of the cell. Surprisingly, however, simulations showed that the correct pattern formed independent of the initial conditions. This finding relied on two conditions. Firstly, like in the ParABS system, uniform binding across the nucleoid and subsequent on-nucleoid diffusion, means that self-organised MukBEF foci can be influenced by the incoming fluxes from either side of the nucleoid. In the case of one off-centre focus, this results in the focus being 'fed' more on one side than the other and thus it moves to the middle of the cell where the incoming fluxes balance. Secondly, robust positioning required stochastic effects, without which foci could become stuck at the poles. This configuration is not stable to stochastic effects and the system is easily pushed to the stable regularly positioned configuration. Together, balancing fluxes and stochasticity allows the self-organised MukBEF foci to sense the nucleoid length and/or each other and position themselves appropriately.

## Discussion: theory and experiment, finding an optimal mix for the future

In this review, we have highlighted two fruitful examples of the interplay between theory and experiment in *E. coli*: cell division positioning and DNA/protein partitioning. Both cases implement centre-finding, but do so in very different ways, utilising spatiotemporal oscillations and flux balance, respectively. For both, mathematical modelling building on substantial experimental foundations was critical for establishing the underlying mechanisms. In such cases where the mechanism involved exceeds our ability as scientists to understand intuitively, the injection of mathematical modelling can provide a solid framework for dealing with complex interactions and can greatly accelerate our acquisition of mechanistic understanding. For the case of the Min system, one can reasonably ask how long it would have taken to get at the heart of the mechanism without input from theoreticians. It is not implausible to think that the scientific community would still be waiting for an answer nearly two decades later.

Despite these successes, it is fair to say that mathematical approaches are still under-represented in the literature in biology. Specifically, within microbiology, this is unfortunate as the comparative simplicity of prokaryotes (complex but not too complex) makes them attractive targets for combined modelling and experiments. To assess the current degree of "market penetration" of modelling within microbiology, we performed literature searches within the ISI Web of Science database, examining the fraction of published papers in experimental microbiology with a modelling component from 2002 to 2017, the most recent year with complete data (Figure 3). We are specifically interested here in published papers that combine both types of investigation, as we believe that such a combination is optimal, certainly as compared to pure theory papers existing in isolation without experimental input, their ideas mostly unread by experimentalists. We therefore searched for microbiology papers with a theoretical component in otherwise experimental journals. We examined first papers in Molecular Microbiology and Journal of Bacteriology, two of the highest impact journals focused specifically on fundamental microbiology. Searching on relevant keywords, we found that the proportion of papers involving modelling increased ~3 fold between 2002 and 2017, but from a low base (from around 0.5%) and reaching only 1-3% in recent years. Searching specifically for bacterial papers in the general journal PNAS did suggest more theory involvement, with the proportion of papers doubling from around 3-5% in the early 2000s to around 7-9% in more recent years. Interestingly, this data suggests that high impact microbiology papers do involve significantly more theoretical content. However, the proportion of papers is still low at less than 10%. The proportion of papers with theoretical content that is optimal for the most rapid scientific progress is, self-evidently, hard to estimate. The two topics reviewed in this publication clearly benefited from theoretical input, but other topics may of course be more straightforward conceptually. Nevertheless, it seems overwhelmingly likely that the optimal proportion is much higher than the 1-2% from the specialist literature, perhaps by an order of magnitude or more, and also probably higher than that seen in PNAS, although by a smaller factor of perhaps 2-3. From this analysis it is hard to escape the conclusion that in microbiology (and biology in general) progress is being hampered by a lack of theoretical thinking.

Overall, we believe that such thinking is best input through collaborations between theoretical and experimental groups. Of course, it is possible for both styles to be combined within a single group and there are successful examples of such work. However, it is often difficult for experimental group leaders to acquire the experience to critically manage theoretical work, which can differ markedly from project to project, and which does not lend itself to the "pipeline" approaches so common in bioinformatics. Similarly, many theorists may struggle with experimental approaches due to lack of training and experience. Given that most microbiologist are not experts in theory, and are not likely to become so, then to optimally assist this cohort of researchers (which constitutes the vast majority of active microbiologists today), collaboration with dedicated theory groups would seem to be the best overall strategy.

Nearly 20 years ago when one of us (MH) started thinking about biology, and how theoretical approaches could be used, it seemed clear that the field was ripe for a step change in the use of theoretical approaches. This still seems true today. Much progress has been made in the intervening years, but in our opinion biology in general, and microbiology in particular, remain very far from utilising the optimum combination of theory and experiment.

# Legends

**Figure 1**

Outline model of the Min system dynamics. A circled P indicates the ATP-bound form of MinD. Membrane shown in grey, with cytoplasm above.

**Figure 2**

Flux-balance as a general positioning mechanism.

(a) A plasmid (red) acts as a sink for ParA-ATP (blue line), which binds uniformly over the nucleoid. (Left) With the plasmid off-centre, the flux (or gradient) of ParA-ATP from either side is different. (Right) If the plasmid moves up the Par-ATP gradient then it will be positioned at the nucleoid centre, where the fluxes balance.

(b) Basal unbinding/hydrolysis of ParA-ATP affects positioning. The length-scale $L = \sqrt{2D/k}$, where $D$ is the nucleoid-bound diffusion constant and $k$ is the basal unbinding rate defines the diffusive length-scale of Par-ATP. If $L$ is less than the nucleoid length, then the plasmid does not receive positional information over the entire nucleoid. Here, $L = 0.2$ in units of nucleoid length, otherwise as in (a). This incomplete sensing may contribute to spatiotemporal fluctuations of the plasmid about the centre position.

**Figure 3**

Plot showing the percentage of published bacteriology articles with a computational or mathematical component for three established journals. Results are based on a Web of Science topic search for any of the phrases 'math*', 'computation*' or 'simulation*'. For the multidisciplinary journal Proceedings of the National Academy of Sciences of the United States of America (PNAS), the search was restricted to bacteriology with the additional topic keyword 'bacteria'.

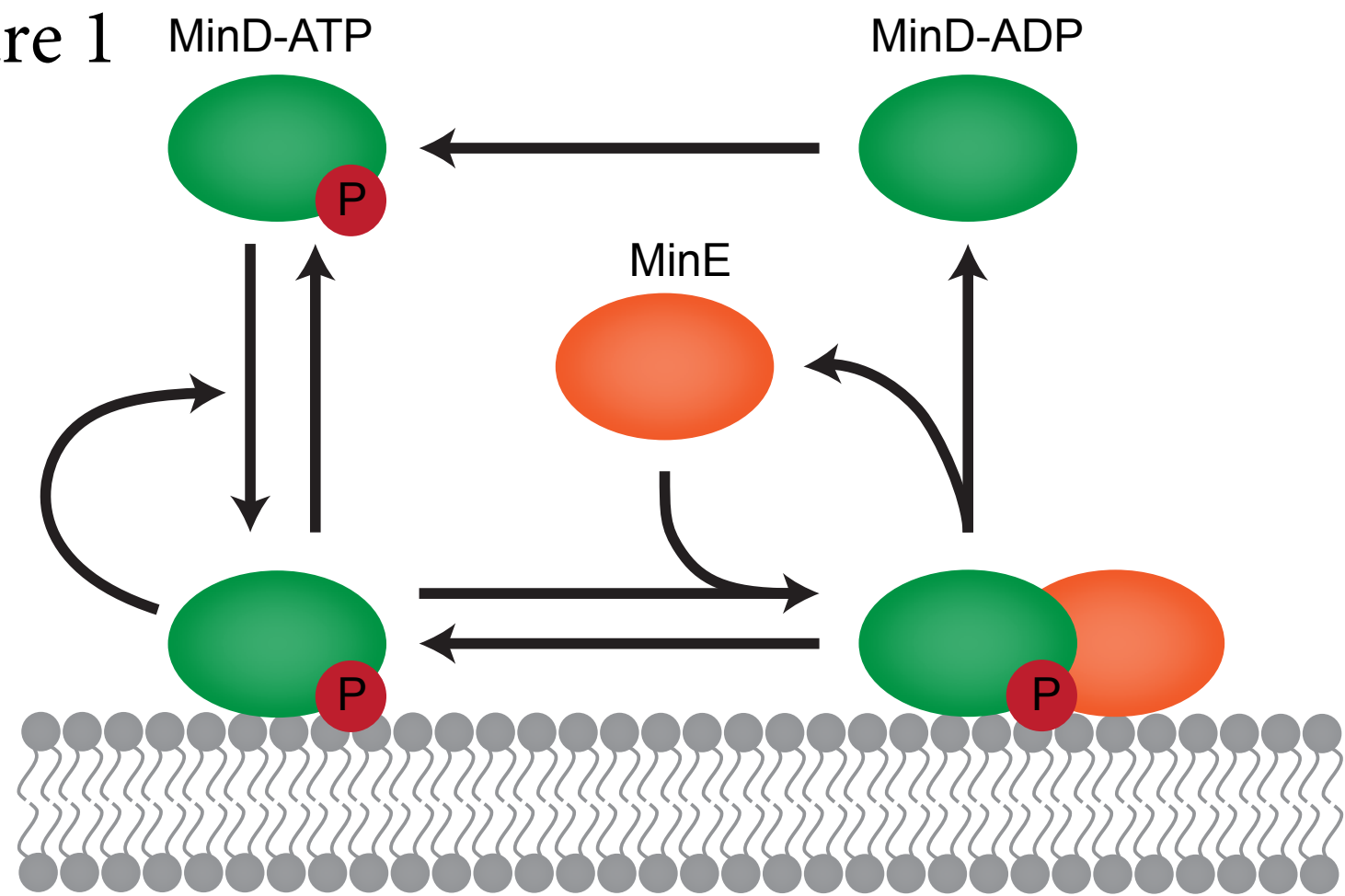

Figure 1

**Figure 2**

**(a)** Unbinding only at the plasmid

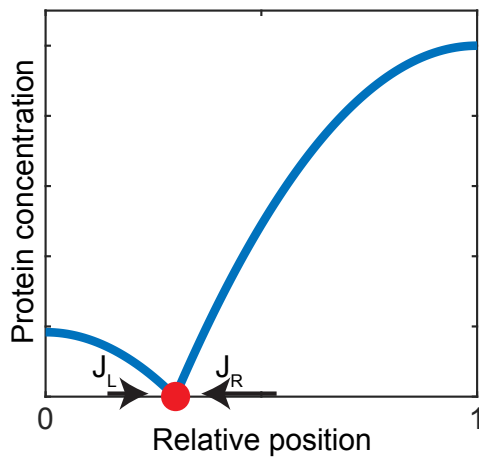
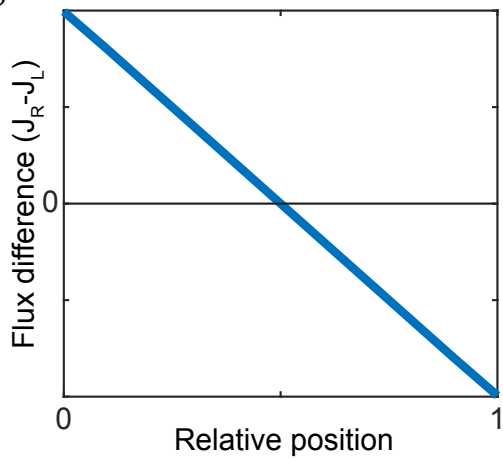

**(b)** With basal unbinding

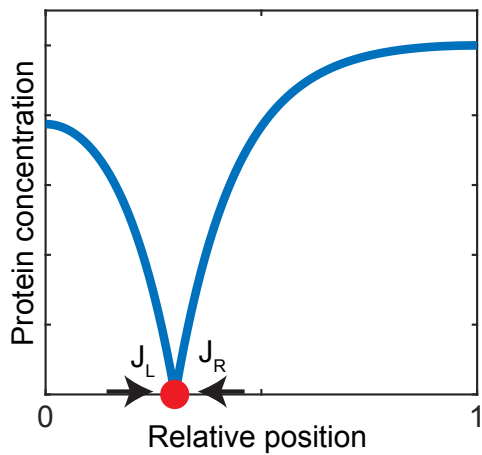
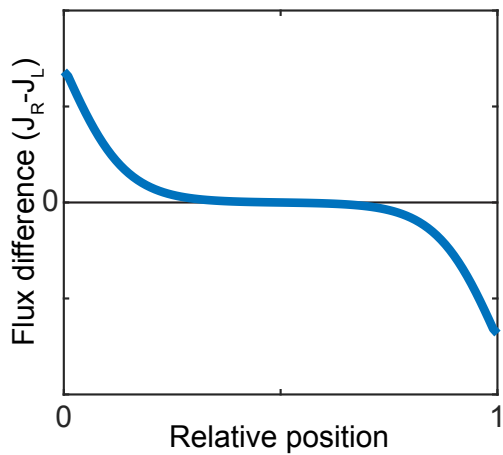

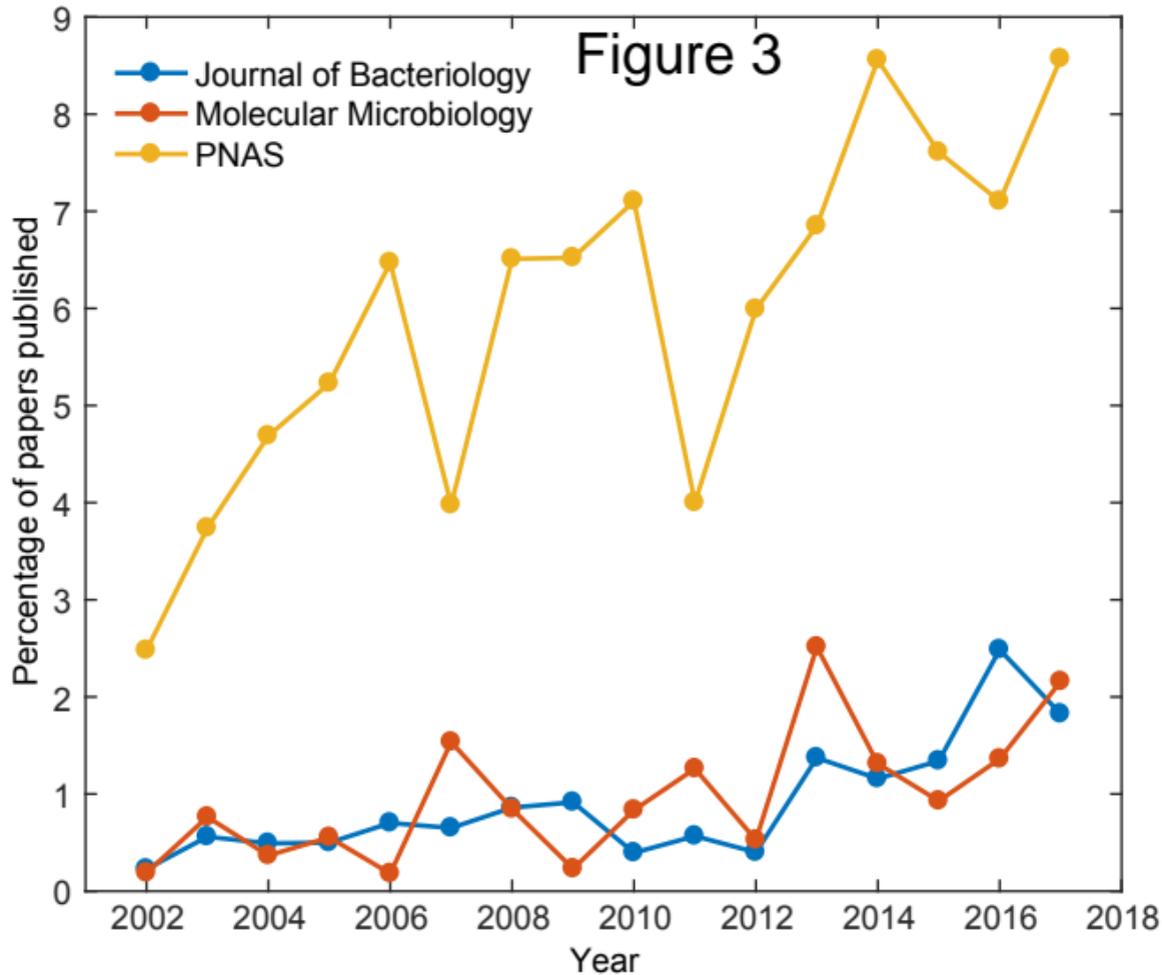

Figure 3